\documentclass[aps,prb,twocolumn,superscriptaddress,a4paper]{revtex4-1}

\usepackage[dvipdfmx]{graphicx}

\usepackage{amsmath,amsthm,amssymb}
\usepackage{amsmath,bm}
\usepackage{color}
\usepackage[svgnames,psnames]{xcolor}
\usepackage{multirow,bigdelim}
\usepackage{physics}

\newcommand{\1}{\mbox{1}\hspace{-0.25em}\mbox{l}}

\begin{document}

\preprint{APS/123-QED}
%%%%%%%%%%
\title{Non-Hermitian $\mathbb{Z}_4$ skin effect protected by glide symmetry}

\author{Sho Ishikawa}
\affiliation{
 Department of Physics, Kyoto University, Kyoto 606-8502, Japan 
}
\author{Tsuneya Yoshida}%
\affiliation{
 Department of Physics, Kyoto University, Kyoto 606-8502, Japan 
}
\affiliation{
 Institute for Theoretical Physics, ETH Zurich, 8093 Zurich, Switzerland
}
%%%%%%%%%%

%%%%%%%%%%
\begin{abstract}
Although nonsymmorphic symmetry protects $\mathbb{Z}_4$ topology for Hermitian systems, non-Hermitian topological phenomena induced by such a unique topological structure remain elusive.
In this paper, we elucidate that systems with glide symmetry exhibit non-Hermitian skin effects (NHSE) characterized by $\mathbb{Z}_4$ topology.
Specifically, numerically analyzing a two-dimensional toy model, we demonstrate that the $\mathbb{Z}_4$ topology induces the NHSE when the topological invariant takes $\nu=1,2$.
Furthermore, our numerical analysis demonstrates that the NHSE is destroyed by perturbations preserving the relevant symmetry when the $\mathbb{Z}_4$-invariant takes $\nu=4$.
\end{abstract}
%%%%%%%%%

\maketitle

%%%%%%%%%%%%%%%%%%
\section{Introduction}
Topology of wavefunctions in insulators and superconductors is one of the central issues in condensed matter physics because of gapless boundary states induced by topology 
in the bulk~\cite{Kane_QSH_PRL2005,Kane_Z2QSH_PRL2005,Bernevig_BHZmodel_Science2006,Qi_TRII_PRB2008,Hasan_TIReview_RMP2010,Qi_TITSC_RMP2011}.
The integer quantum Hall system~\cite{Klitzing_QHR_PRL1980,Halperin_QHC_PRB1982,TKNN_QHC_PRL1982,Avron_Homotopy_PRB1983,Hatsugai_EdgeIQHE_PRB1993,Hatsugai_CQHE_PRL1993,Gusynin_UnconventIQHE_PRL2005} is an example of topological insulators without symmetry where the chiral edge mode results in quantized Hall conductance with high accuracy. 
Topological structures are enriched by time-reversal symmetry, particle-hole symmetry, and chiral symmetry~\cite{Su_SSH_PRB1980,Altland_AZclass_PRB1997,Kitaev_KitaevChain_PU2001,Fu_TRSZ2_PRB2006} as systematically clarified by the 10-fold way classification~\cite{Schnyder_TISC3D_PRB2008,Kitaev_PeriodicTable_ACP2009,Ryu_TITSC_PRB2010,Chiu_TQMS_APS2016};
three (two) symmetry classes allow $\mathbb{Z}$ ($\mathbb{Z}_2$) topology in each case of spatial dimensions.
Topological structures are further enriched by crystalline symmetry~\cite{Teo_BiSb_PRB2008,Fu_TCI_PRL2011,Hsieh_TCI_NC2012,Tanaka_TCI_NatPhys2012,Chiu_TopoReflection_PRB2013,Morimoto_TopoAddSym_PRB2013,Shiozaki_TopoSG_PRB2014}.
In particular, unique topological structures may exist under nonsymmorphic symmetry~\cite{Liu_Nonsymmorphic_PRB2014,Shiozaki_Z2Mobius_PRB2015,Fang_Nonsymmorphic_PRB2015,Shiozaki_TopoNSG_PRB2016,Wang_Hourglass_Nature2016,Schoop_DiracNonsymmorphic_NC2016,Zhao_Nonsymmorphic_PRB2016,Kruthoff_HermiNonsymm_PRX2017,Ma_ExperiNonsymmor_SA2017,Zhang_Nonsymmorphic_PRL2023,Max_1DZ4_arXiv2024,Daido_Z4Moebius_PRL2019,Chang_MobiusKondo_NatPhys2017}.
For instance, $\mathbb{Z}_4$ topology is allowed under glide symmetry which is described by a product of reflection and the half-translation. 
This unique $\mathbb{Z}_4$ topology results in distinctive surface states analogous to a M\"obius strip~\cite{Shiozaki_TopoNSG_PRB2016}. 

The notion of topology is further extended to non-Hermitian systems, which elucidated that non-Hermiticity induces novel topological phenomena without Hermitian counterparts~\cite{Hatano_Nelson_PRL1996,Hatano_Nelson_PRB1997,Hu_nH_PRB11,Esaki_nH_PRB11,TELeePRL16_Half_quantized,Gong_TPNH_PRX2018,Kawabata_NHSymTopo_PRX2019,Ashida_NHphysics_AdvPhys2020,Bergholtz_ExceptNH_RMP2021,Okuma_NHTopoReview_AnnRev2023,Lin_TopoNHSE_FrontPhys2023,Kunst_BulkBoundary_PRL2018,Yokomizo_BBC_PRL19,Shen_nH_PRL2018,Kozii_nH_PRB17,Yoshida_EP_DMFT_PRB18}.
A representative example is the non-Hermitian skin effect (NHSE) which is the extreme sensitivity of eigenvalues and eigenstates to the boundary conditions due to point-gap topology~\cite{ Yao_EdgeTopoNH_PRL2018,Yao_NHChern_PRL2018,Lee_AnatomySkin_PRB2019,Okuma_TopoOrigin_PRL2020,Borgnia_NHBoundary_PRL2020,Zhang_WindingSkin_PRL2020,Zhang_NHSE_AnnPhysX2022}.
Because of non-trivial point-gap topology, most of all eigenmodes are localized around only one of the edges which are known as skin modes~\cite{Okuma_TopoOrigin_PRL2020}.
While the above NHSE is reported for systems with no symmetry, symmetry-protected NHSEs are also reported for systems with time-reversal symmetry~\cite{Okuma_TopoOrigin_PRL2020} or reflection symmetry~\cite{Yoshida_MirrorSkin_PRResearch2020}.
The NHSEs are reported for a variety of classical systems
~\cite{Hofmann_ReciprocalSkin_PRResearch2020,Helbig_NHTopoCircuit_NatPhys2020,Shuo_NHElecCircuit_Research2021,Zou_NHTopoCircuit_NatComm2021,Zhang_HighNHSE_NatComm2021,Shang_NHSEMachineLearning_AdvSci2022,Zhang_AcousticNHSE_NatComm2021,Zhang_ElecCircuit_PRB2023,
Weidemann_TopoFunneling_Sci2020,Song_NHSEPhotonic_PRApp2020,Xiao_PTsymExcept_PRL2021,Wang_DetectNonBloch_PRL2021,
Brandenbourger_NonReciprocal_NatComm2019,Ghatak_NHTopo_PNAS2020,Wang_NHTopo_SciAdv2023}
as well as quantum systems~\cite{Gou_NonrecQuantTrans_PRL2020,Xiao_NHQuantDyn_NatPhys2020,Palacios_ActiveParticles_NatComm2021,Liang_NHSESignatures_PRL2022,Shen_NHSEFermiSkin_arXiv2023}.

The above two progresses imply the potential presence of exotic non-Hermitian phenomena under nonsymmorphic symmetry.
Although several works explored the non-Hermitian topology under nonsymmorphic symmetry~\cite{Wu_APTNHSSH_PRB2021,Tanaka_NHSENonsymmor_PRB2024,Koenig_NodalNHWallpaper_APL2024}, non-Hermitian topological phenomena induced by $\mathbb{Z}_4$ topology remain elusive.

In this paper, we numerically demonstrate the emergence of the NHSE induced by $\mathbb{Z}_4$ topology in systems with glide symmetry.
Specifically, we elucidate that the NHSE emerges in a two-dimensional toy model when the $\mathbb{Z}_4$ topological invariant takes $\nu = 1,2$.
The NHSE characterized by $\nu = 2$ appears only for boundaries where glide symmetry is closed. In addition, one-dimensional topology is trivial for the toy model of $\nu = 2$.
Furthermore, we observe that the NHSE is destroyed by perturbations preserving the relevant symmetry for $\nu = 4$, which indicates that the NHSE is induced by $\mathbb{Z}_4$ topology.
The above results elucidate the emergence of the non-Hermitian $\mathbb{Z}_4$ skin effect protected by glide symmetry.

The rest of this paper is organized as follows.
In Sec.~\ref{sec:Z4inv_sym}, we discuss the $\mathbb{Z}_4$-invariant in non-Hermitian systems with glide symmetry.
In Sec.~\ref{sec:NHSE_Z4}, we observe the emergence of NHSE characterized by the $\mathbb{Z}_4$-invariant.
Section~\ref{sec:summary} provides a brief summary of our work.
Appendices are devoted to symmetry of doubled Hermitian Hamiltonians, their edge modes, and one-dimensional topology. 
%%%%%%%%%%%%%%%%%%

%%%%%%%%%%%%%%%%%%
\section{$\mathbb{Z}_4$-topological invariant and relevant symmetry}
\label{sec:Z4inv_sym}
%%%%%%%%%%%%%%%%%%
We consider a two-dimensional non-Hermitian Hamiltonian with the following symmetry ($\mathrm{TRS}^\dagger$):
\begin{equation}
    \label{eq:TRS^dagger}
    \mathcal{T} H^T(\bm{k}) \mathcal{T}^{-1} = H(-\bm{k}),
\end{equation}
where $\mathcal{T}$ is a unitary matrix satisfying $\mathcal{T}\mathcal{T}^*=-1$.
This non-Hermitian Hamiltonian belongs to~\cite{Kawabata_NHSymTopo_PRX2019} class $\mathrm{AII}^\dagger$. 
Furthermore, we add glide symmetry for the non-Hermitian system:
\begin{equation}
    \label{eq:NH_glide_symmetry}
    \mathcal{G}(k_x) H(\bm{k}) \mathcal{G}^{-1}(k_x) = H^\dagger(\bm{k}),
\end{equation}
where $\mathcal{G}(k_x)$ is a glide operator satisfying $\mathcal{G}^2(k_x) = -e^{-ik_x}$ and $\mathcal{T}\mathcal{G}(k_x)\mathcal{T}^{-1} = \mathcal{G}^\dagger(k_x)$.

In the presence of symmetry constraints [Eqs.~(\ref{eq:TRS^dagger})~and~(\ref{eq:NH_glide_symmetry})], the non-Hermitian topology is characterized by a $\mathbb{Z}_4$-topological invariant.
In order to see this, we consider the following doubled Hermitian Hamiltonian with a reference energy $E_{\mathrm{ref}}\in \mathbb{C}$:
\begin{equation}
    \label{eq:doubled_Hamiltonian}
    \tilde{H}(E_{\mathrm{ref}}) =
    \left(\begin{array}{cc}
        0 & H - E_{\mathrm{ref}} \1 \\
        H^\dagger - E^*_{\mathrm{ref}}\1 & 0
    \end{array}\right),
\end{equation}
where $\1$ is the identity matrix.
This Hermitian Hamiltonian belongs to class DIII and has glide symmetry (for more details see Appendix~\ref{appsub:DH_sym}).
The above Hamiltonian $\tilde{H}(E_\mathrm{ref})$ is known to be characterized by the following $\mathbb{Z}_4$-invariant~\cite{Shiozaki_TopoNSG_PRB2016} $\nu$:
\begin{align}
    \label{eq:Hermitian_DIII_Z_4}
    \nu \equiv & \frac{2i}{\pi} \int_{-\pi}^\pi {d}k_y \operatorname{tr} \mathcal{A}^{I}_{+} (k_x = \pi, k_y) \notag\\
    &-\frac{i}{\pi} \int_0^\pi {d}k_x \int_{-\pi}^\pi {d}k_y \operatorname{tr} \mathcal{F}_{+} (k_x, k_y) \quad (\mathrm{mod}~4),
\end{align}
where $\mathcal{A}_{+}$ and $\mathcal{F}_{+}$ are respectively the Berry connection and the Berry curvature for occupied states of $\tilde{H}(E_{\mathrm{ref}})$ in the plus-eigensector of the glide operator $G(k_x)$ [for the explicit definition of $G(k_x)$ see Eq.~(\ref{eq:glide_operator}) in Appendix~\ref{appsub:DH_sym}] with eigenvalues $g_\pm(k_x) = \pm ie^{-ik_x/2}$.
The superscript $I$ denotes one of the Kramers pair on the one-dimensional subspace specified by $k_x = \pi$ where the time-reversal operation is closed for the plus-eigensector.

%%%%%%%%%%%%%%%%%%
\section{Non-Hermitian skin effect characterized by $\mathbb{Z}_4$-topological invariant}
\label{sec:NHSE_Z4}
%%%%%%%%%%%%%%%%%%
We numerically elucidate the emergence of the NHSE induced by $\mathbb{Z}_4$ topology under glide symmetry. 
Specifically, our numerical analysis demonstrates the emergence of an NHSE for $\nu=1,2$. 
Such an NHSE is destroyed by perturbations preserving the relevant symmetry for $\nu=4$.

%%%%%%%%%%%%%%%%%%
\subsection{NHSE for $\nu = 1$}
\label{subsec:NHSE_nu1}
\begin{figure}[t]
    \begin{minipage}{0.48\linewidth}
        \includegraphics[width=0.95\linewidth]{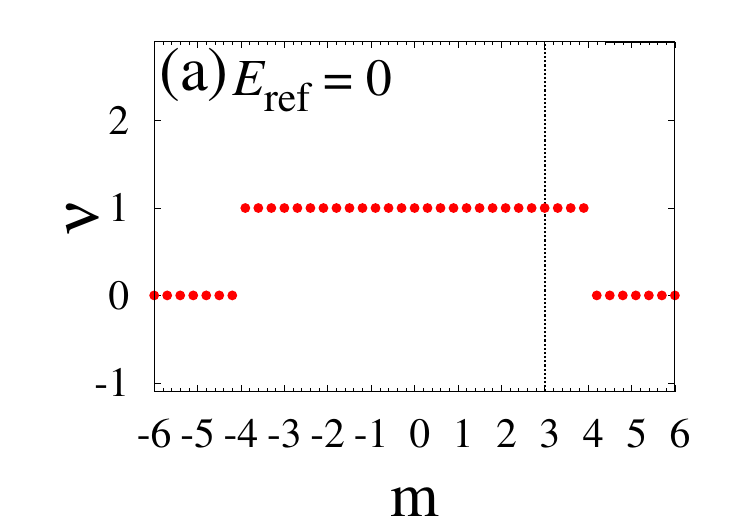} 
    \end{minipage}
    \begin{minipage}{0.48\linewidth}
        \includegraphics[width=0.95\linewidth]{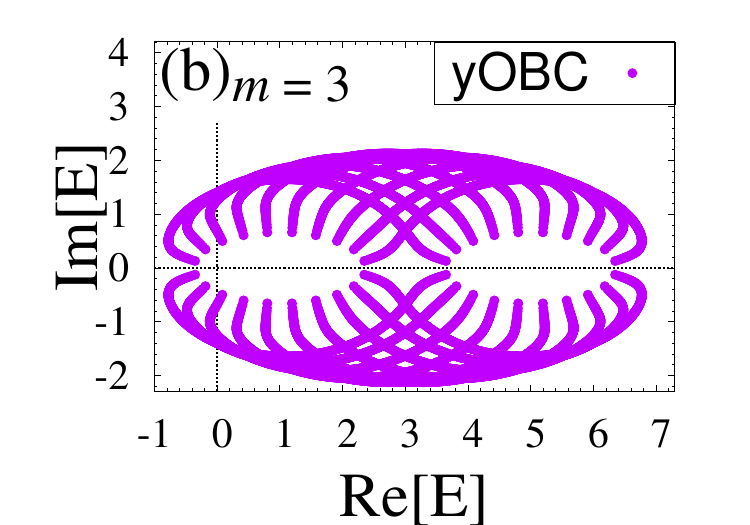}
    \end{minipage}  
    \begin{minipage}{0.48\linewidth}
         \includegraphics[width=0.95\linewidth]{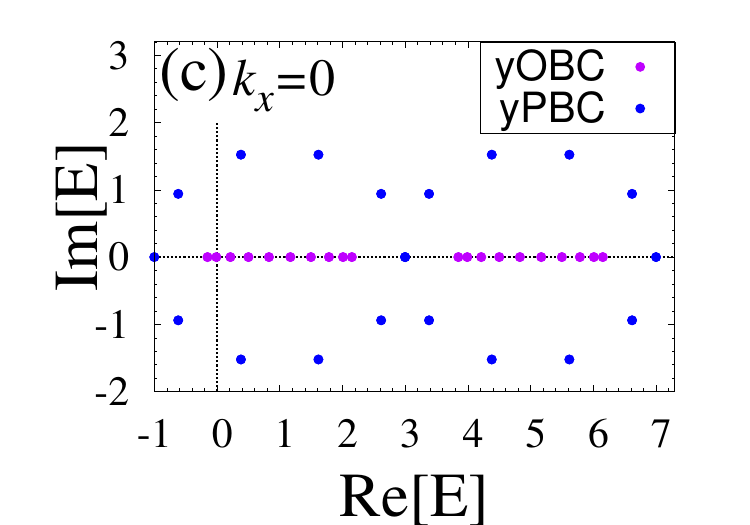}
    \end{minipage}
    \begin{minipage}{0.48\linewidth}             
        \includegraphics[width=0.95\linewidth]{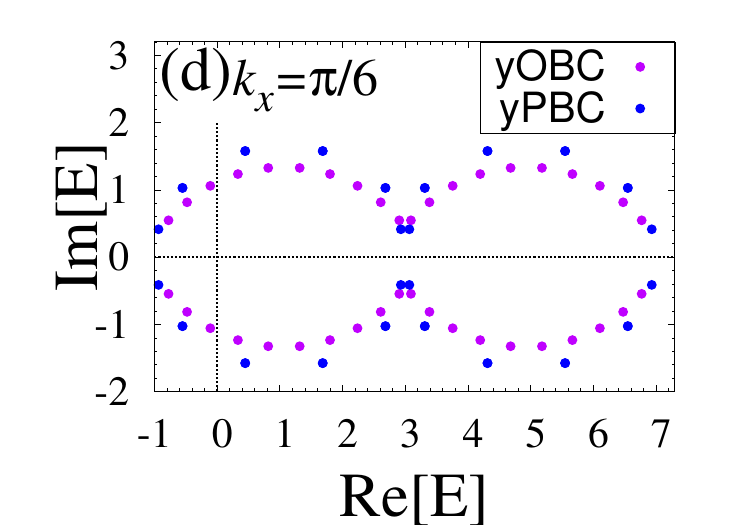}
    \end{minipage}
    \caption{ (a): The $\mathbb{Z}_4$-invariant of $H_1$ [Eq.~(\ref{eq:nu=1_model})] as a function of $m$.
     The vertical dotted line denotes $m=3$.
     (b): Energy spectra of $H_1$ [Eq.~(\ref{eq:nu=1_model})] under yOBC for $k_x = 2\pi n/10^3~(n = 1, \ldots, 10^3-1)$. The more detailed data obtained for $k_x = 2\pi n/10^4~(n = 1, \ldots, 10^4-1)$ are essentially the same as panel (b).
     (c)[(d)]: The energy spectra for $m=3$ and $k_x=0$ [$k_x=\pi/6$].
     Data denoted by purple (blue) dots are obtained under yOBC (yPBC).
     These data are obtained for $(L, t, t_{\mathrm{sp}}) = (10, 1, 0.8)$.}
    \label{fig:nu1_invariant_spectra}
\end{figure}

%%%%%%%%%%%%%%%%%%
To investigate the system for $\nu=1$, we consider the following Hamiltonian:
\begin{eqnarray}
    \label{eq:nu=1_model}
        &&H_1(k_x, k_y) \nonumber \\  
        &&\quad \equiv \left[ m\eta_0 - t\left\{ \left(1+\cos{k_x}\right)\eta_x -\sin{k_x}\eta_y\right\}\right] \rho_0 \nonumber \\  
        &&\quad \quad + i t_{\mathrm{sp}}\left[\sin{k_x}\eta_x - (1-\cos{k_x})\eta_y\right]\rho_z \nonumber \\ 
        &&\quad \quad - i 2t_{\mathrm{sp}}\sin{k_y}\eta_0\rho_y.
\end{eqnarray}
Here, $m, t, t_{\mathrm{sp}},$ and $\alpha$ are real numbers. 
Pauli matrices are denoted by $\eta_i$ and $\rho_i$ $(i=x,y,z)$.
Employing a numerical method of the discretized Brillouin zone~\cite{Fukui_ChernNum_JPSJ2005,Fukui_Z2Num_JPSJ2007,Yoshida_Z4Num_PRB2019}, we can see that the $\mathbb{Z}_4$-invariant takes $\nu=1$ for $-4<m<4$ and $(L, t, t_{\mathrm{sp}}) = (10, 1, 0.8)$ [see Fig.~\ref{fig:nu1_invariant_spectra}(a)].

We numerically analyze the energy spectrum and right eigenstates by changing boundary conditions.
By xOBC (yOBC), we denote open boundary conditions in the $x$- ($y$-) direction.
By xPBC (yPBC), we denote periodic boundary conditions in the $x$- ($y$-) direction.
We set the parameter $m$ to $3$ where the $\mathbb{Z}_4$-invariant takes $\nu=1$ [see Fig.~\ref{fig:nu1_invariant_spectra}(a)].
Unless otherwise noted, periodic boundary conditions are imposed in the $x$-direction (i.e., xPBC).

The energy spectra are plotted in Figs.~\ref{fig:nu1_invariant_spectra}(b)-\ref{fig:nu1_invariant_spectra}(d).
For $k_x = 0$, the spectrum under yOBC is real while the spectrum under yPBC becomes complex.
However, for $k_x \neq 0$, the spectrum becomes complex for both cases of yOBC and yPBC.

\begin{figure}[htbp]
    \begin{minipage}{0.48\linewidth}
        \includegraphics[width=0.95\linewidth]{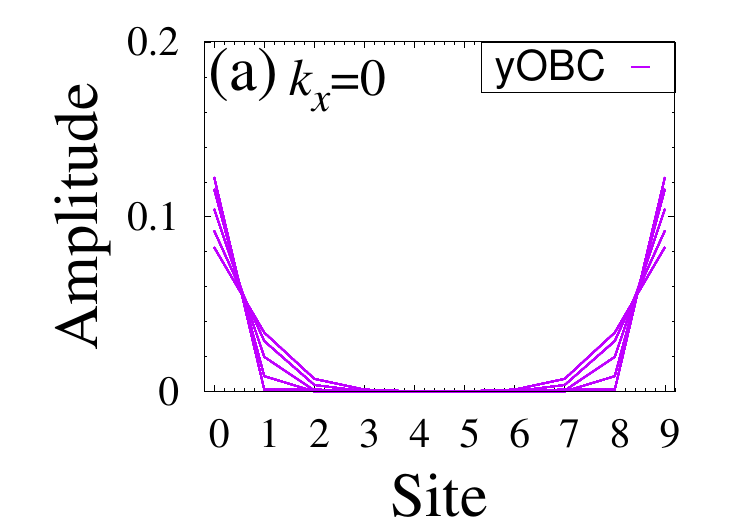}
    \end{minipage}
    \begin{minipage}{0.48\linewidth}
        \includegraphics[width=0.95\linewidth]{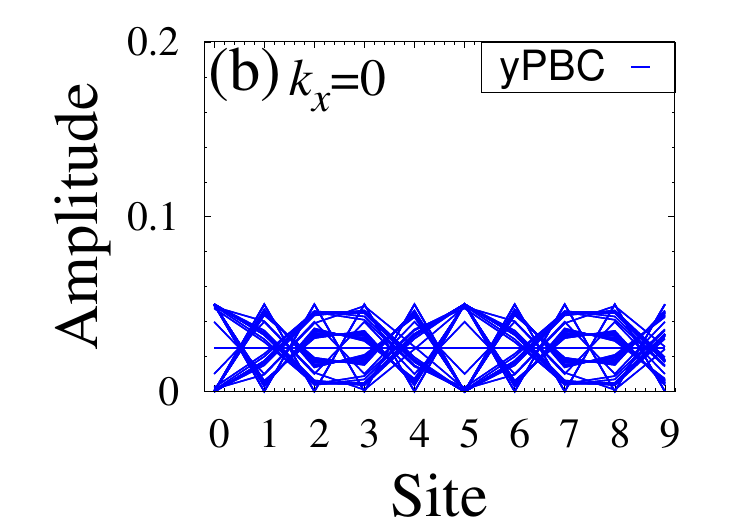}
    \end{minipage}
    \begin{minipage}{0.48\linewidth}
        \includegraphics[width=0.95\linewidth]{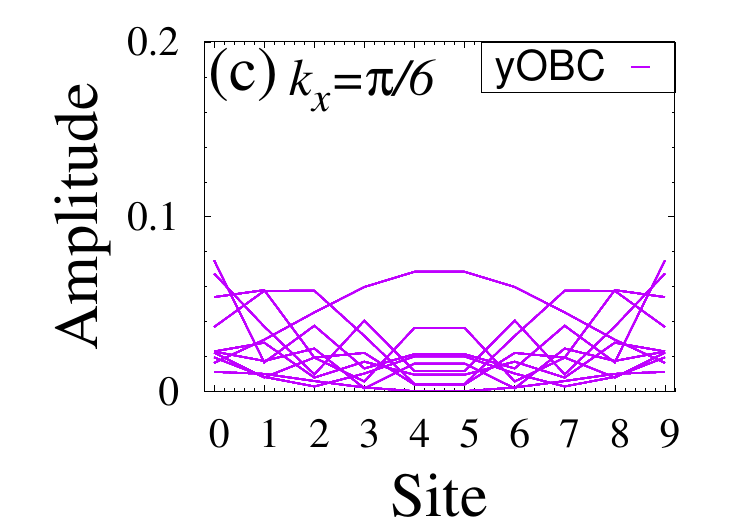}
    \end{minipage}
    \begin{minipage}{0.48\linewidth}
        \includegraphics[width=0.95\linewidth]{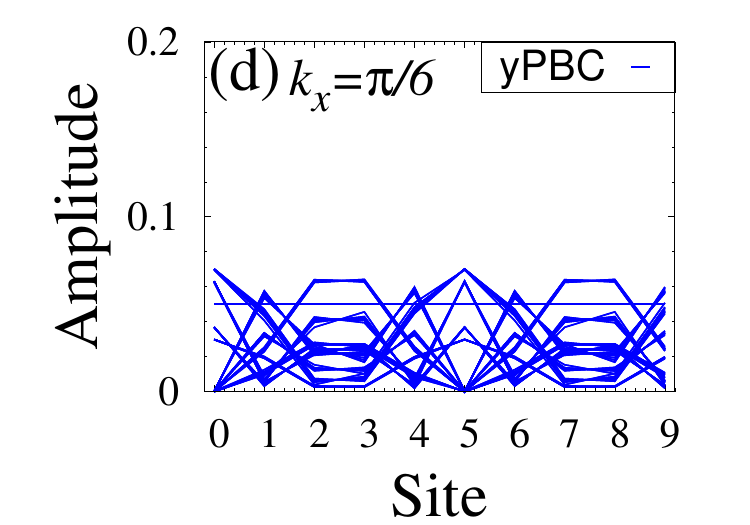}
    \end{minipage}
    \caption{ (a)[(b)]: Amplitude of right eigenstates of $H_1$ [see Eq.~(\ref{eq:nu=1_model})] under yOBC [yPBC] for $k_x=0$.
    (c)[(d)]: Amplitude of the right eigenstates under yOBC [yPBC] for $k_x=\pi/6$.
    The amplitude is defined as $\abs{\braket{i_y}{\Psi_{nR}}}^2$ where $\ket{\Psi_{nR}}$ are the right eigenstates of the Hamiltonian (i.e., $H_1\ket{\Psi_{nR}}=E_n\ket{\Psi_{nR}}$).
    These data are obtained for $(L, m, t, t_{\mathrm{sp}}) = (10, 3, 1, 0.8)$.
    We have introduced a perturbation~\cite{TRSB_ptb_ftnt} [Eq.~(\ref{eq:nu=1_model_ptb})] with $\beta = 10^{-12}$.
    }
    \label{fig:NHSE_eigen_nu1}
\end{figure}

The right eigenstates are plotted in Fig.~\ref{fig:NHSE_eigen_nu1}.
For $k_x = 0$, skin modes emerge around both edges.
Namely, eigenstates under yOBC are localized at both ends of the system [Fig.~\ref{fig:NHSE_eigen_nu1}(a)] while eigenstates under yPBC are delocalized [Fig.~\ref{fig:NHSE_eigen_nu1}(b)].
However, for $k_x \neq 0$, eigenstates under yOBC are delocalized, which is similar to the case of yPBC [see Figs.~\ref{fig:NHSE_eigen_nu1}(c)~and~\ref{fig:NHSE_eigen_nu1}(d)].

To obtain these data, a perturbation~\cite{TRSB_ptb_ftnt}
\begin{equation}
    \label{eq:nu=1_model_ptb}
    H_{1,\mathrm{\mathrm{ptb}}} = -i\beta \eta_0 \rho_x,
\end{equation}
with a small real number $\beta$ is added to lift degeneracy by breaking $\mathrm{TRS}^\dagger$.
With the above results, we can conclude that the NHSE is observed for the Hamiltonian defined in Eq.~(\ref{eq:nu=1_model}) where the $\mathbb{Z}_4$-invariant takes $\nu=1$.

%%%%%%%%%%%%
%%%%%%%%%%%%

%%%%%%%%%%%%
%%%%%%%%%%%%

%%%%%%%%%%%%
\subsection{NHSE for $\nu = 2$}
\label{subsec:NHSE_nu2}
\begin{figure}[htbp]
    \begin{minipage}{0.48\linewidth}
        \includegraphics[width=0.95\linewidth]{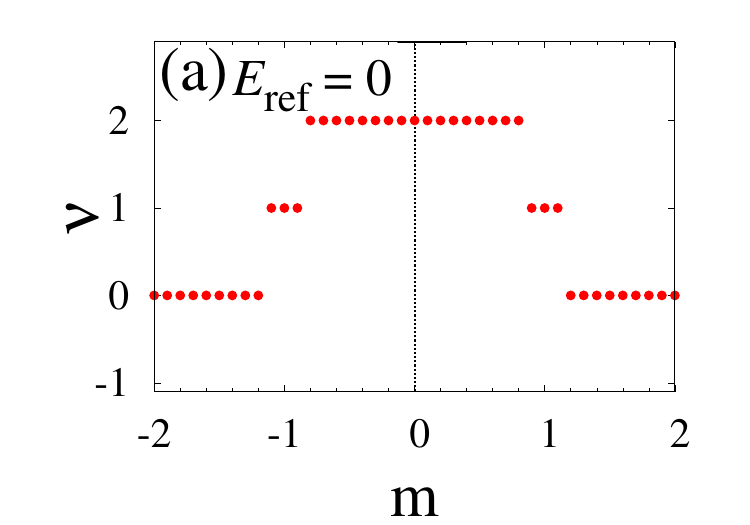}
    \end{minipage}
    \begin{minipage}{0.48\linewidth}
        \includegraphics[width=0.95\linewidth]{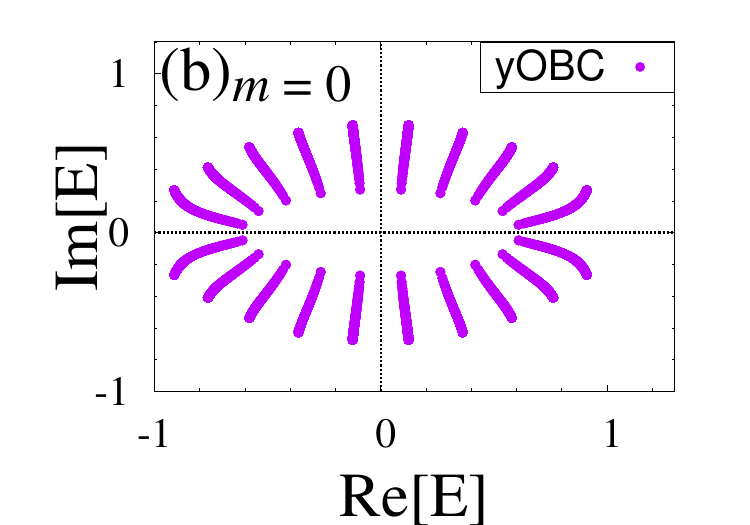}
    \end{minipage}
    \begin{minipage}{0.48\linewidth}
        \includegraphics[width=0.95\linewidth]{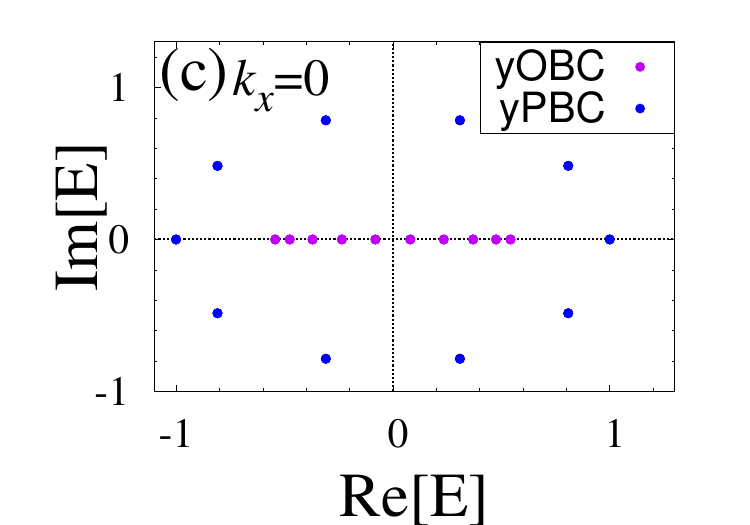}
    \end{minipage}
    \begin{minipage}{0.48\linewidth}
        \includegraphics[width=0.95\linewidth]{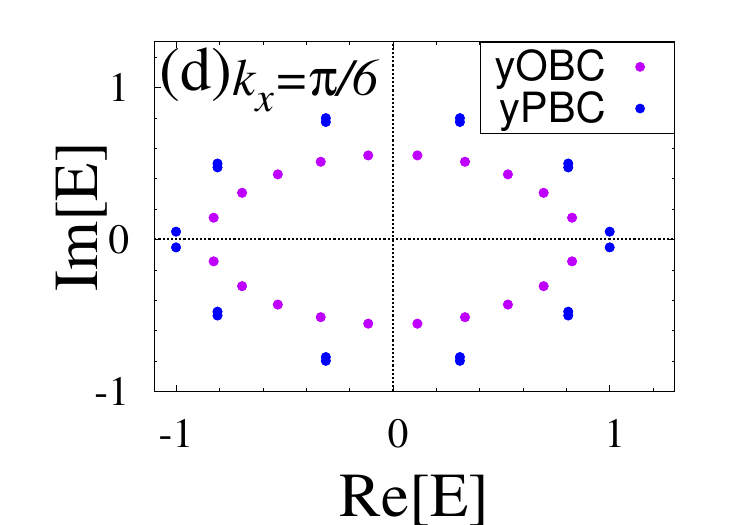}
    \end{minipage}\\
    \caption{ (a): The $\mathbb{Z}_4$-invariant of $H_2$ [Eq.~(\ref{eq:nu=2_model})] as a function of $m$.
     The vertical dotted line denotes $m=0$.
     (b): Energy spectra of $H_2$ [Eq.~(\ref{eq:nu=2_model})] under yOBC for $k_x = 2\pi n/10^3~(n = 1, \ldots, 10^3-1)$. 
     The more detailed data obtained for $k_x = 2\pi n/10^4~(n = 1, \ldots, 10^4-1)$ are essentially the same as panel (b).
     (c)[(d)]: The energy spectra for $m=0$ and $k_x=0$ [$k_x=\pi/6$].
     Data denoted by purple (blue) dots are obtained under yOBC (yPBC).
     These data are obtained for $(L, t, \kappa, \Delta, \alpha) = (10, 1, 0.2, 0.8, 0.1)$.}
    \label{fig:nu2_invariant_spectra}
\end{figure}

%%%%%%%%%%%%
To investigate the system for $\nu=2$, we consider the following Hamiltonian:
\begin{eqnarray}
    \label{eq:nu=2_model}
    &&H_2(k_x, k_y) \nonumber\\
    &&\quad \equiv \left( m - t\cos{k_y} \right) \eta_0\rho_0 \nonumber\\
    &&\quad \quad + i \, \kappa \sin{(k_x/2)} \left[\cos{\left(k_x/2\right)} \eta_x + \sin{\left(k_x/2\right)} \eta_y\right] \rho_y \nonumber\\
    &&\quad \quad+ i \Delta \sin{k_y} \eta_0\rho_z - i2\alpha \sin{k_y} \eta_z\rho_x.
\end{eqnarray}
Here, $\kappa$ and $\Delta$ are real numbers. 
The $\mathbb{Z}_4$-invariant takes $\nu=2$ for $-0.8 \leq m \leq 0.8$ and $(L, t, \kappa, \Delta, \alpha) = (10, 1, 0.2, 0.8, 0.1)$ [see Fig. \ref{fig:nu2_invariant_spectra}(a)].

We numerically analyze the energy spectrum and right eigenstates in a similar way to the previous case.
We set the parameter $m$ to $0$ where the $\mathbb{Z}_4$-invariant takes $\nu=2$ [see Fig.~\ref{fig:nu2_invariant_spectra}(a)].

The energy spectra are plotted in Figs.~\ref{fig:nu2_invariant_spectra}(b)-\ref{fig:nu2_invariant_spectra}(d).
For $k_x = 0$, the spectrum under yOBC is real while the spectrum under yPBC becomes complex.
However, for $k_x \neq 0$, the spectrum becomes complex for both cases of yOBC and yPBC.

\begin{figure}[htbp]
    \begin{minipage}{0.48\linewidth}
        \includegraphics[width=0.95\linewidth]{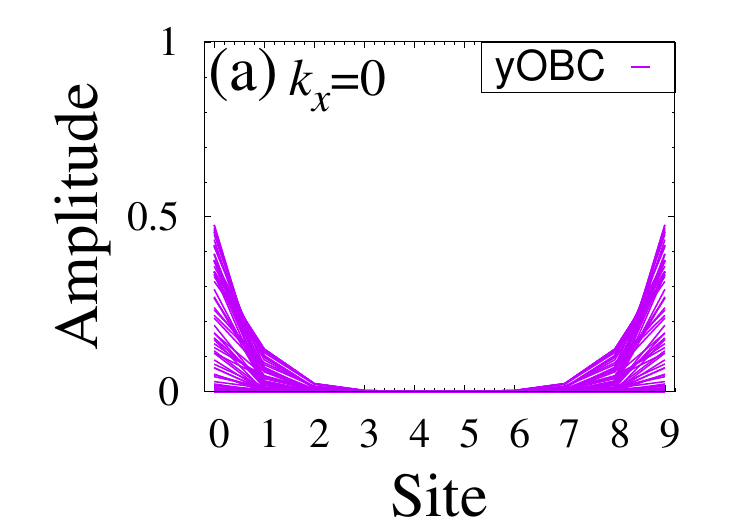}
    \end{minipage}
    \begin{minipage}{0.48\linewidth}
        \includegraphics[width=0.95\linewidth]{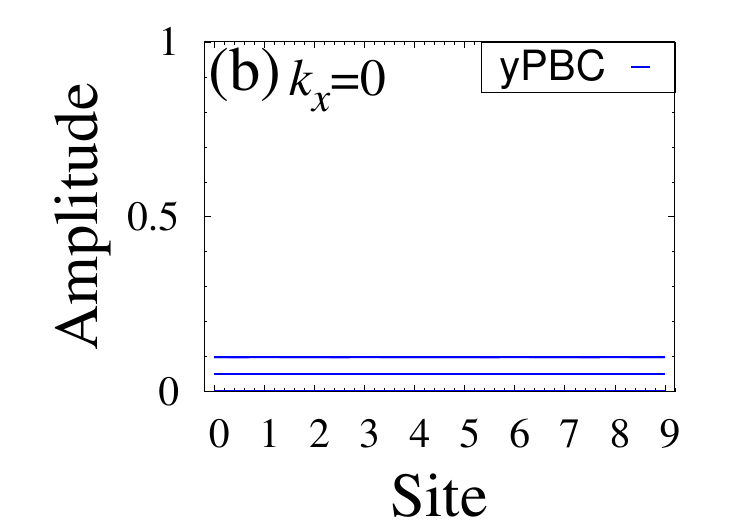}
    \end{minipage}
    \begin{minipage}{0.48\linewidth}
        \includegraphics[width=0.95\linewidth]{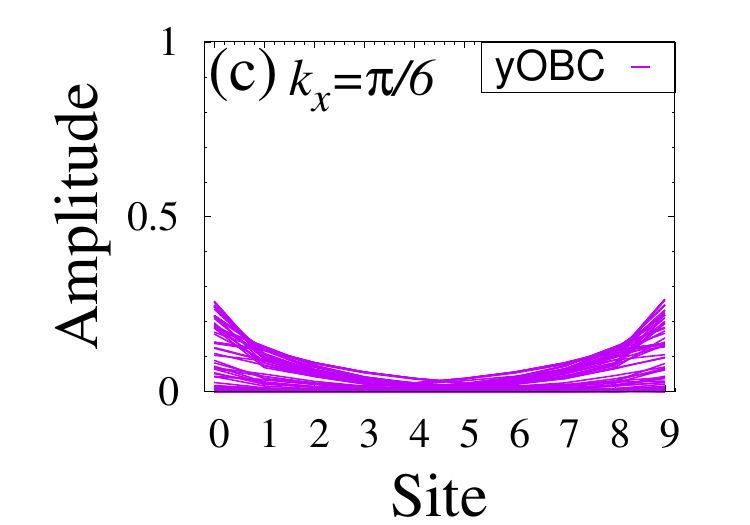}
    \end{minipage}
    \begin{minipage}{0.48\linewidth}
        \includegraphics[width=0.95\linewidth]{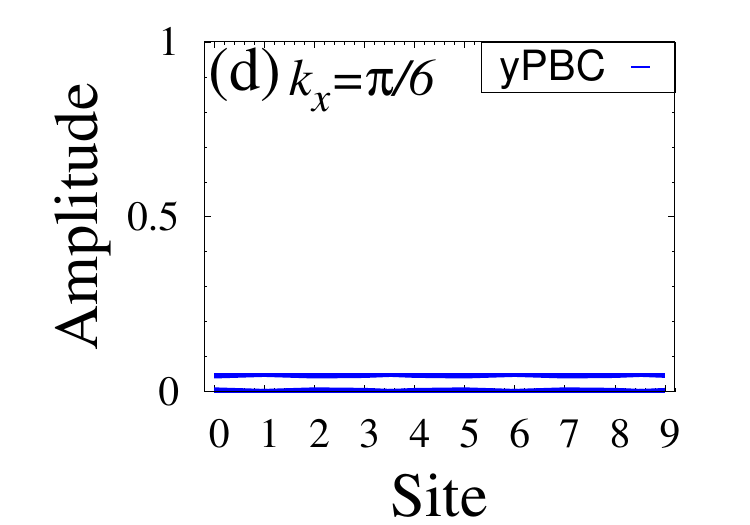}
    \end{minipage}
    \caption{ (a)[(b)]: Amplitude of right eigenstates of $H_2$ [see Eq.~(\ref{eq:nu=2_model})] under yOBC [yPBC] for $k_x=0$. 
    (c)[(d)]: Amplitude of the right eigen states under yOBC [yPBC] for $k_x=\pi/6$. 
    The amplitude is defined as $\abs{\braket{i_y}{\Psi_{nR}}}^2$ where $\ket{\Psi_{nR}}$ are the right eigenstates (i.e., $H_2\ket{\Psi_{nR}}=E_n\ket{\Psi_{nR}}$).
    These data are obtained for $(L, m, t, \kappa, \Delta, \alpha) = (10, 0, 1, 0.2, 0.8, 0.1)$.
    We have introduced a perturbation~\cite{TRSB_ptb_ftnt} [Eq.~(\ref{eq:nu=2_model_ptb})] with $\beta = 10^{-12}$.
    }
    \label{fig:NHSE_eigen_nu2}
\end{figure}

The right eigenstates are plotted in Fig.~\ref{fig:NHSE_eigen_nu2}.
For $k_x = 0$, skin modes emerge around both edges.
Namely, eigenstates under yOBC are localized at both ends of the system [Fig.~\ref{fig:NHSE_eigen_nu2}(a)] while eigenstates under yPBC are delocalized [Fig.~\ref{fig:NHSE_eigen_nu2}(b)].
In contrast, for $k_x \neq 0$, eigenstates under yOBC are delocalized, which is similar to the case of yPBC [see Figs.~\ref{fig:NHSE_eigen_nu2}(c)~and~\ref{fig:NHSE_eigen_nu2}(d)].
To obtain these data, a perturbation~\cite{TRSB_ptb_ftnt}
\begin{equation}
    \label{eq:nu=2_model_ptb}
    H_{2,\mathrm{\mathrm{ptb}}} = -i\beta \eta_0 \rho_x,
\end{equation}
with a small real number $\beta$ is added to lift degeneracy by breaking $\mathrm{TRS}^\dagger$.

With the above results, we can conclude that the NHSE is observed for Hamiltonian defined in Eq.~(\ref{eq:nu=2_model}) where the $\mathbb{Z}_4$-invariant takes $\nu = 2$.

Here, one may consider that the above NHSE is induced by one-dimensional topology of class AII${}^\dagger$.
We note, however, that the corresponding topological invariant takes zero for this model (see Appendix~\ref{sec:topo_without_glide}).
We also note that the NHSE disappears when we impose xOBC and yPBC where the glide symmetry is not closed (see Appendix~\ref{sec:xOBC_yPBC}).

%%%%%%%%%%%%
%%%%%%%%%%%%

%%%%%%%%%%%%
%%%%%%%%%%%%

%%%%%%%%%%%%
\subsection{NHSE destroyed by perturbations for $\nu=4$}
\label{subsec:NHSE_nu4}
%%%%%%%%%%%%
Stacking two copies of $H_2$ [Eq.~(\ref{eq:nu=2_model})] yields a bilayer system where the $\mathbb{Z}_4$-invariant takes $\nu=4$.
Introducing a coupling between the layers destroys the NHSE with preserving the relevant symmetry, which demonstrates that the NHSE observed in Sec.~\ref{subsec:NHSE_nu2} is induced by $\mathbb{Z}_4$ topology.

\begin{figure}[htbp]
        \begin{minipage}{0.48\linewidth}
            \includegraphics[width=0.95\linewidth]{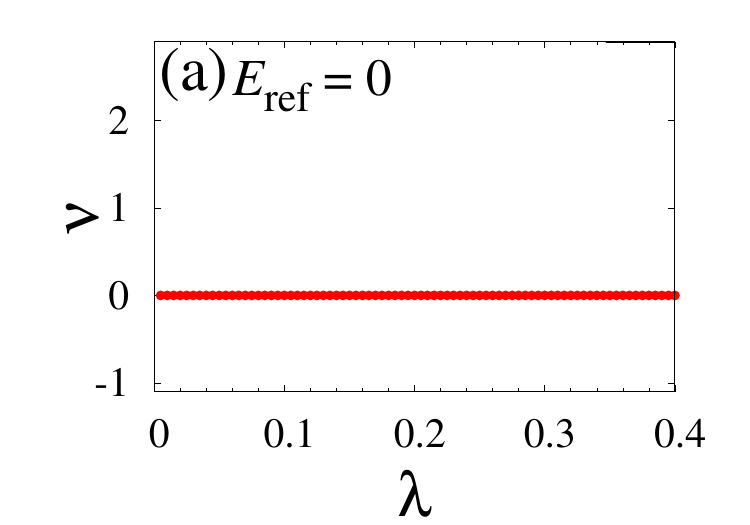}
        \end{minipage}
        \begin{minipage}{0.48\linewidth}
            \includegraphics[width=0.95\linewidth]{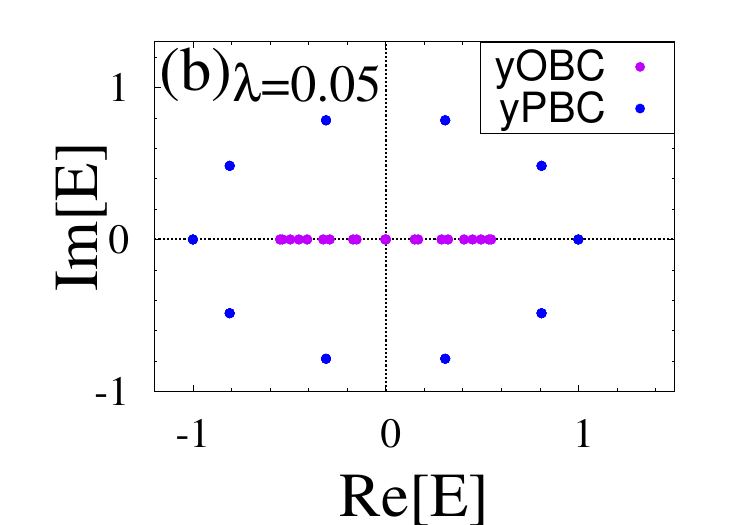}
        \end{minipage}
        \begin{minipage}{0.48\linewidth}
            \includegraphics[width=0.95\linewidth]{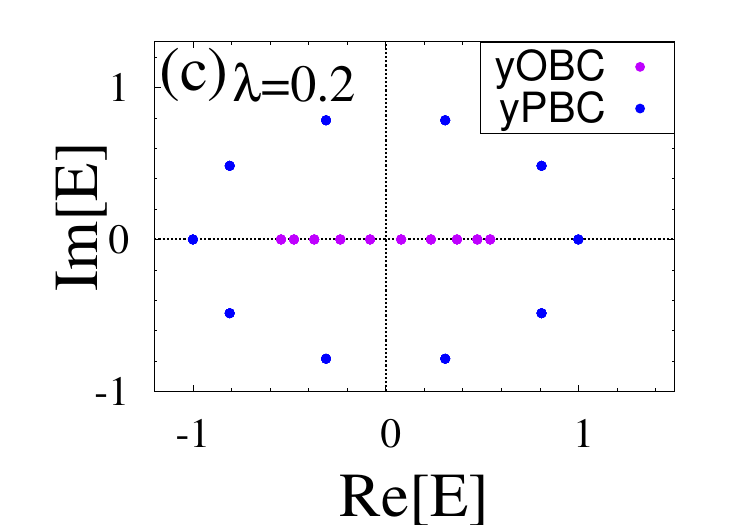}
        \end{minipage}
        \begin{minipage}{0.48\linewidth}
            \includegraphics[width=0.95\linewidth]{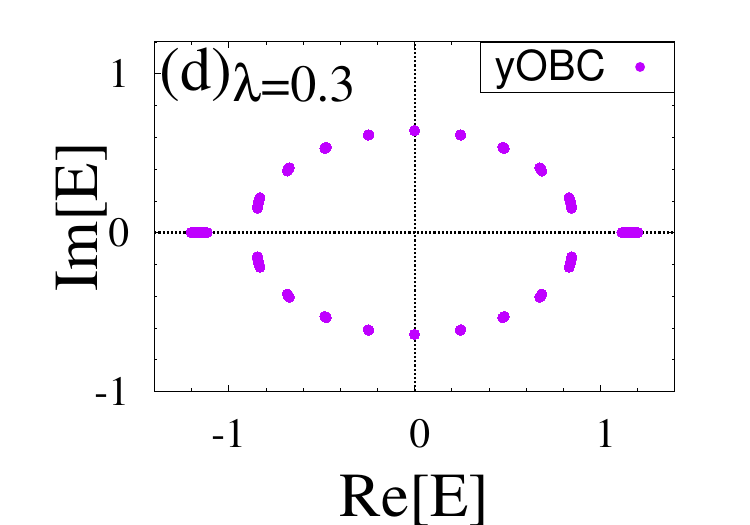}
        \end{minipage}\\
    \caption{ (a): The $\mathbb{Z}_4$-invariant of $H_4$ [Eq.~(\ref{eq:nu=4_model})] as a function of $\lambda$.
    (b)[(c)]: Energy spectra of $H_4$ [Eq.~(\ref{eq:nu=4_model})] for $\lambda=0.05$ [$0.2$] and $k_x=0.50536050426$ [$\pi$].
    Data denoted by purple (blue) dots are obtained under yOBC (yPBC).
    (d): The energy spectra under yOBC for $k_x = 2\pi n/10^3~(n = 0, \ldots, 10^3-1)$. The more detailed data obtained for $k_x = 2\pi n/10^4~(n = 0, \ldots, 10^4-1)$ are essentially the same as panel (d).
    These data are obtained for $(L, m, t, \kappa, \Delta, \alpha) = (10, 0, 1, 0.2, 0.8, 0.1)$.
    }
    \label{fig:nu4_invariant_spectra}
\end{figure}

The above destruction of the NHSE is observed in the following Hamiltonian:
\begin{equation}
    \label{eq:nu=4_model}
    H_4(k_x, k_y) = H_2(k_x, k_y)\zeta_0 + \lambda \eta_0 \rho_x \zeta_y.
\end{equation}
Here, $\lambda$ is a real positive number. 
Pauli matrices are denoted by $\zeta_i$ $(i=x,y,z)$.
The second term preserves glide symmetry, $\mathcal{G}(k_x) \lambda \eta_0\rho_x\zeta_y \mathcal{G}^{-1}(k_x) =\lambda \eta_0\rho_x\zeta_y$.
As shown in Fig. \ref{fig:nu4_invariant_spectra}(a), the $\mathbb{Z}_4$-invariant takes $\nu=0$ for this model.

The energy spectrum is plotted in Figs.~\ref{fig:nu4_invariant_spectra}(b)-\ref{fig:nu4_invariant_spectra}(d).
For $\lambda = 0.05$ and $0.2$, the spectrum under yOBC is real while the spectra under yPBC become complex when $k_x$ takes a specific value.
However, for $\lambda = 0.3$, the sensitivity of eigenvalues to boundary conditions is not observed;
the spectrum under yOBC becomes complex for arbitrary $k_x$, which is similar to the case of yPBC.

\begin{figure}[htbp]
    \begin{minipage}{0.48\linewidth}
        \includegraphics[width=0.95\linewidth]{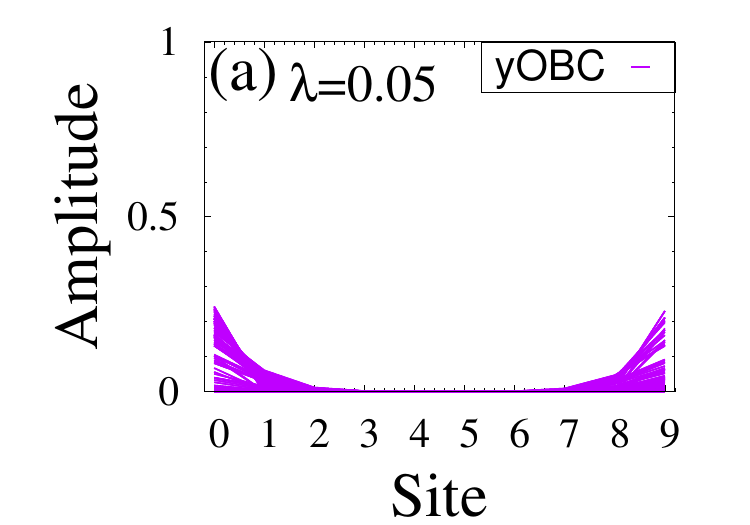}
    \end{minipage}
    \begin{minipage}{0.48\linewidth}
        \includegraphics[width=0.95\linewidth]{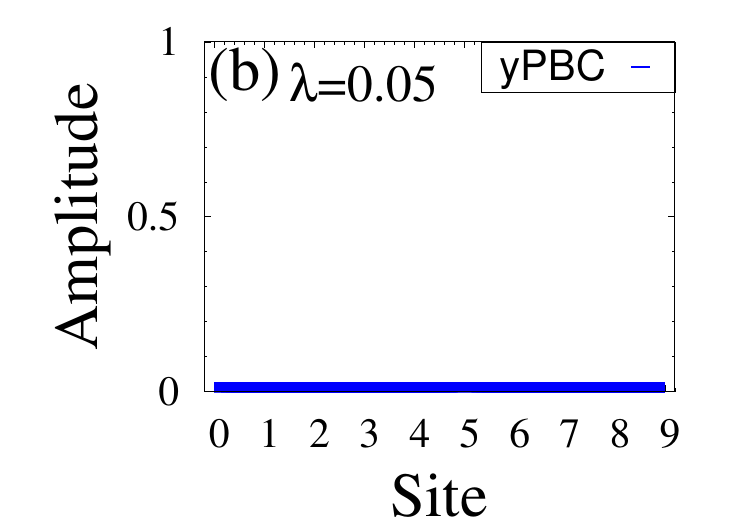}
    \end{minipage}
    \begin{minipage}{0.48\linewidth}
        \includegraphics[width=0.95\linewidth]{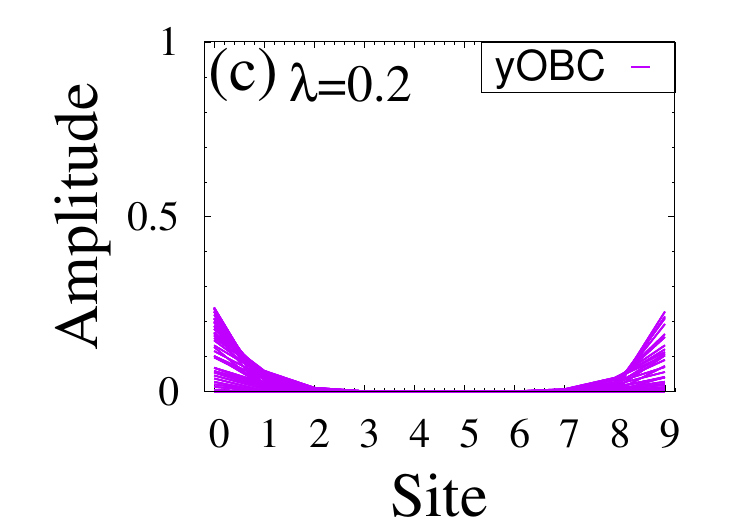}
    \end{minipage}
    \begin{minipage}{0.48\linewidth}
        \includegraphics[width=0.95\linewidth]{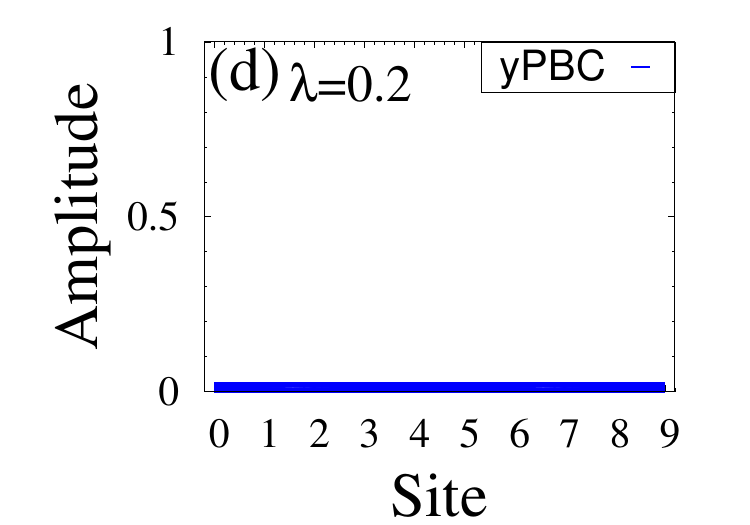}
    \end{minipage}
    \begin{minipage}{0.48\linewidth}
        \includegraphics[width=0.95\linewidth]{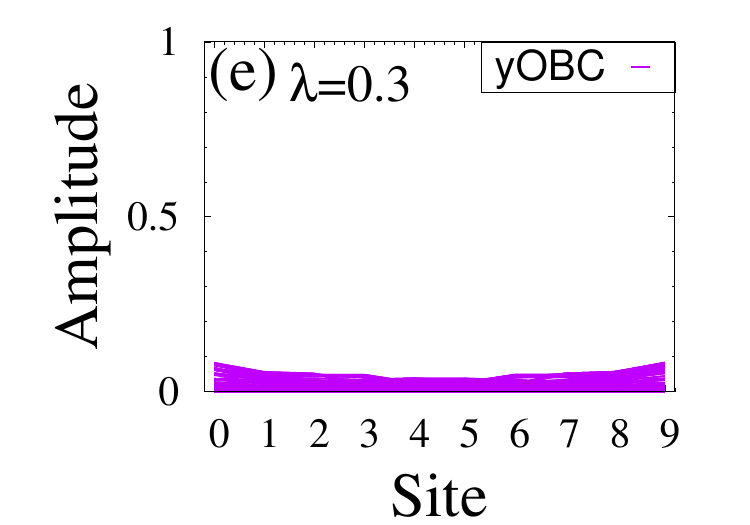}
    \end{minipage}
    \begin{minipage}{0.48\linewidth}
        \includegraphics[width=0.95\linewidth]{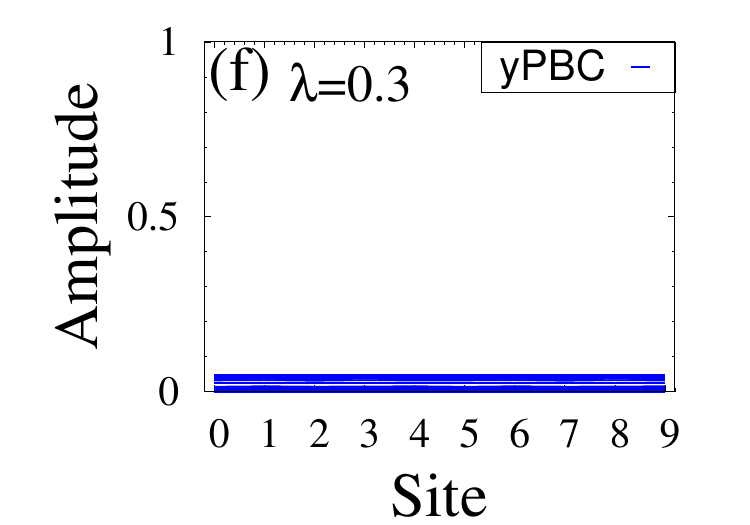}
    \end{minipage}\\
    \caption{ (a)[(b)]: Amplitude of right eigenstates of $H_4$ [see Eq.~(\ref{eq:nu=4_model})] under yOBC [yPBC] for $\lambda=0.05$ and $k_x=0.50536050426$.
    (c)[(d)]: Amplitude of the right eigenstates under yOBC [yPBC] for $\lambda=0.2$ and $k_x=\pi$.
    (e)[(f)]: Amplitude of the right eigenstates under yOBC [yPBC] for $\lambda=0.3$ and $k_x=\pi/6$. 
    For $\lambda=0.3$, the data obtained for other wave numbers are essentially the same as panel (e)[(f)]. 
    Amplitude is defined as $\abs{\braket{i_y}{\Psi_{nR}}}^2$ where $\ket{\Psi_{nR}}$ are right eigenstates (i.e., $H_4\ket{\Psi_{nR}}=E_n\ket{\Psi_{nR}}$).
    These data are obtained for $(L, m, t, \kappa, \Delta, \alpha) = (10, 0, 1, 0.2, 0.8, 0.1)$.
    We have introduced a perturbation~\cite{TRSB_ptb_ftnt} [Eq.~(\ref{eq:nu=4_model_ptb})] with $\beta = 10^{-12}$.
    }
    \label{fig:NHSE_eigen_nu4}
\end{figure}

The right eigenstates are plotted in Fig.~\ref{fig:NHSE_eigen_nu4}.
For $\lambda = 0.05$ and $0.2$, the eigenstates under yOBC are localized at both ends of the system at a specific value of $k_x$ [see Figs.~\ref{fig:NHSE_eigen_nu4}(a) and \ref{fig:NHSE_eigen_nu4}(c)] while the eigenstates under yPBC are delocalized [see Figs.~\ref{fig:NHSE_eigen_nu4}(b) and \ref{fig:NHSE_eigen_nu4}(d)].
However, for $\lambda=0.3$, the above localized modes are not observed;
the eigenstates under yOBC are delocalized, which is similar to the case of PBC [see Figs.~\ref{fig:NHSE_eigen_nu4}(e)~and~\ref{fig:NHSE_eigen_nu4}(f)].
To obtain these data, a perturbation~\cite{TRSB_ptb_ftnt}
\begin{equation}
    \label{eq:nu=4_model_ptb}
    H_{4,\mathrm{\mathrm{ptb}}} = -i\beta \eta_0 \rho_x \zeta_0,
\end{equation} 
with a small real number $\beta$ is added to lift degeneracy by breaking $\mathrm{TRS}^\dagger$.

The above results demonstrate that when the $\mathbb{Z}_4$-invariant takes $\nu=4$, the NHSE is destroyed by perturbations preserving the time-reversal symmetry and glide symmetry.
This fact supports that $\mathbb{Z}_4$ topology induces the NHSEs observed for Hamiltonians defined in Eqs.~(\ref{eq:nu=1_model})~and~(\ref{eq:nu=2_model}).

In this section, we have numerically demonstrated that systems with $\nu=1,2$ exhibit NHSEs while the NHSE is destroyed by perturbations preserving the relevant symmetry for $\nu=4$.
We consider that the NHSE emerges for a system with $\nu=3$ because flipping the sign of the third term of Eq.~(\ref{eq:nu=1_model}) [$-i2t_{\mathrm{sp}}\sin{k_y} \to i2t_{\mathrm{sp}}\sin{k_y}$] yields the system with $\nu=-1$ [$=3 \ (\mathrm{mod}4)$] which exhibits NHSE.
Putting the above results together, we end up with the emergence of non-Hermitian $\mathbb{Z}_4$ skin effects protected by glide symmetry.

%%%%%%%%%%%%
%%%%%%%%%%%%

%%%%%%%%%%%%
%%%%%%%%%%%%

\newpage
%%%%%%%%%%%%%%%%%%%%%%%%
\section{Summary}
\label{sec:summary}
%%%%%%%%%%%%%%%%%%%%%%%%
In this paper, we have elucidated that $\mathbb{Z}_4$ topology with glide symmetry induces NHSEs.
Specifically, we have numerically analyzed two-dimensional toy models characterized by the $\mathbb{Z}_4$-invariant $\nu$ under glide symmetry.
Our numerical analysis has clarified that systems with $\nu=1,2$ exhibit the NHSE.
The NHSE for $\nu=2$ is observed only on boundaries where the glide symmetry is closed.
In addition, the one-dimensional subsystem at $k_x=0$ is trivial for the toy model of $\nu=2$ [Eq.~(\ref{eq:nu=2_model})].
Furthermore, we have observed that the NHSE characterized by $\nu=4$ is destroyed by perturbations preserving the relevant symmetry.
The above results indicate the emergence of the non-Hermitian $\mathbb{Z}_4$ skin effect protected by glide symmetry.

In Hermitian systems, it is known that glide symmetry results in M\"obius surface states.
Effects of glide symmetry on skin modes are left as a future work to be addressed.

%%%%%%%%%%%%%%%%%%%%%%%%
\begin{acknowledgments}
The authors thank Manfred Sigrist, Shu Hamanaka, and Tsugumi Matsumoto for fruitful discussions.
This work is supported by JSPS KAKENHI Grant Nos.~JP21K13850 and JP23KK0247, JSPS Bilateral Program No.~JPJSBP120249925.
T.Y is grateful for the support from the ETH Pauli Center for Theoretical Studies and the Grant from Yamada Science Foundation.
\end{acknowledgments}
%%%%%%%%%%%%%%%%%%%%%%%%

%apsrev4-2.bst 2019-01-14 (MD) hand-edited version of apsrev4-1.bst
%Control: key (0)
%Control: author (8) initials jnrlst
%Control: editor formatted (1) identically to author
%Control: production of article title (0) allowed
%Control: page (0) single
%Control: year (1) truncated
%Control: production of eprint (0) enabled
%

%\newpage
%%%%%%%%%%%%%%%%%%

\appendix

%\newpage.
%%%%%%%%%%%%%%%%%%
\section{$\mathbb{Z}_4$ topology in Hermitian systems}
\label{app:Z4_Hermi}
%%%%%%%%%%%%%%%%%%

%%%%%%%%%%%%%%%%%%
\subsection{Symmetry constraints on Hermitian Hamiltonians}
\label{appsub:DH_sym}
%%%%%%%%%%%%%%%%%%
The symmetry class of Hermitian Hamiltonian defined in Eq.~(\ref{eq:doubled_Hamiltonian}) is the class DIII because it preserves the time-reversal symmetry (TRS) with $\Theta$, the particle-hole symmetry (PHS) with $C$, and the chiral symmetry (CS) with $\Gamma$;
\begin{eqnarray}
    \Theta \tilde{H}(\bm{k}) \Theta^{-1} &=& \tilde{H}(-\bm{k}), \label{eq:TRS}\\
    C \tilde{H}(\bm{k}) C^{-1} &=& -\tilde{H}(-\bm{k}), \label{eq:PHS}\\
    \Gamma \tilde{H}(\bm{k}) \Gamma^{-1} &=& -\tilde{H}(\bm{k}). \label{eq:CS}
\end{eqnarray}
Here, each of these symmetry operators is represented as
$\Theta = \left(\begin{array}{cc}
    0 & \mathcal{T} \\
    \mathcal{T} & 0
\end{array}\right)\mathcal{K}$, $C = \left(\begin{array}{cc}
    0 & \mathcal{T} \\
    -\mathcal{T} & 0
\end{array}\right)\mathcal{K}$, and $\Gamma = \left(\begin{array}{cc}
    \1 & 0 \\
    0 & -\1
\end{array}\right)$. 
Operator $\mathcal{K}$ takes complex conjugation. 
The glide symmetry is written as
\begin{equation}
    \label{eq:H_glide_symmetry}
    G(k_x) \tilde{H}(\bm{k}) G^{-1}(k_x) = \tilde{H}(\bm{k}),
\end{equation}
with
\begin{equation}
    \label{eq:glide_operator}
    G(k_x) = \left(\begin{array}{cc}
        0 & \mathcal{G}(k_x) \\
        \mathcal{G}(k_x) & 0
    \end{array}\right),
\end{equation}
satisfying $\left\{G(k_x), \Theta C\right\} = 0$.

%%%%%%%%%%%%%%%%%%
\subsection{Hermitian Hamiltonians of $\nu=1,2,4$ and non-Hermitian toy models [Eqs.~(\ref{eq:nu=1_model}),~(\ref{eq:nu=2_model}),~and~(\ref{eq:nu=4_model})]}
\label{app:toy_model_Hermi}
%%%%%%%%%%%%%%%%%%
The doubled Hermitian Hamiltonian of $\nu=1$ for the toy model defined in Eq.~(\ref{eq:nu=1_model}) is given by imposing glide symmetry to the BHZ model \cite{Bernevig_BHZmodel_Science2006,Shiozaki_TopoNSG_PRB2016}:
\begin{align}
    \label{eq:BHZ_model}
    H_\mathrm{BHZ}(k_x, k_y) =& \lbrack m - 2t(\cos{k_x}+\cos{k_y})\rbrack \tau_z s_0 \notag\\
    & + 2t_{\mathrm{sp}}\sin{k_x}\tau_y s_0 \notag\\
    & - 2t_{\mathrm{sp}}\sin{k_y}\tau_x s_z.
\end{align}
Here, $m, t,$ and $t_{\mathrm{sp}}$ are real numbers.
Pauli matrices are denoted by $\tau_i$ and $s_i$ $(i=x,y,z)$;
$\tau$'s act on the orbital space and $s$'s act on the spin space.
Adding a staggered modulation of the lattice in the $x$-direction and performing the unitary transformation, we obtain
\begin{align}
    \label{eq:BHZ_glide_model}
    &\tilde{H}_1(k_x, k_y) \notag\\
    & \quad = \lbrack (m-2t\cos{k_y})\eta_0 - t\{ (1 + \cos{k_x})\eta_x-\sin{k_x}\eta_y\} \rbrack \tau_z s_0 \notag\\
    & \quad\quad + t_{\mathrm{sp}}\left[ \sin{k_x}\eta_x - (1-\cos{k_x})\eta_y\right] \tau_y s_0 \notag\\
    & \quad\quad - 2t_{\mathrm{sp}}\sin{k_y}\eta_0\tau_x s_z,
\end{align}
where $\eta_i~(i=x,y,z)$ are Pauli matrices acting on the two inequivalent sites.
The symmetry class of Hermitian Hamiltonian $\tilde{H}_1$ is class DIII since it preserves TRS [Eq.~(\ref{eq:TRS})] with $\Theta = i\eta_0\tau_0 s_y\mathcal{K}$, PHS [Eq.~(\ref{eq:PHS})] with $C = \eta_0\tau_x s_0\mathcal{K}$, and CS [Eq.~(\ref{eq:CS})] with $\Gamma = \eta_0\tau_x s_y$.
$\tilde{H}_1$ also satisfies the glide symmetry [Eq.~(\ref{eq:H_glide_symmetry})] with
$G(k_x) = i\left(\begin{array}{cc}
    0 & 1 \\
    e^{-ik_x} & 0
\end{array}\right)_\eta \tau_0 s_z$.
If we choose the basis such that CS is represented as $\Gamma = \rho_0\chi_z$ where $\rho$'s and $\chi$'s are Pauli matrices, $\tilde{H}_1$ is written as
\begin{equation}
    \label{eq:DH_nu1_model}
    \tilde{H}_1(k_x, k_y) = \left(\begin{array}{cc}
        0 & H_1(k_x, k_y) \\
        H_1^\dagger(k_x, k_y) & 0
    \end{array}\right)_\chi.
\end{equation}
This Hamiltonian is the doubled Hermitian Hamiltonian with $E_\mathrm{ref} = 0$ for the toy model defined in Eq.~(\ref{eq:nu=1_model}).

The doubled Hermitian Hamiltonian of $\nu=2$ for the toy model [Eq.~(\ref{eq:nu=2_model})] is given by adding TRS to the following Hamiltonian which is constructed from spinless chiral $p$-wave superconductors on the square lattice \cite{Shiozaki_TopoNSG_PRB2016}:
\begin{align}
    \label{eq:DH_nonchiral_model}
    &H_{e_+ - e_-}(k_x, k_y) \notag\\
    & \quad = (m-t\cos{k_y}) \eta_0\tau_z \notag\\
    & \quad\quad + \kappa \sin{(k_x/2)} \lbrack \cos{(k_x/2)} \eta_x + \sin{(k_x/2)} \eta_y \rbrack \tau_x \notag\\
    & \quad\quad + \Delta \sin{k_y} \eta_0\tau_y.
\end{align}
Here, $t, \kappa,$ and $\Delta$ are real numbers.
Pauli matrices are denoted by $\tau_i$ and $\eta_i$ $(i = x,y,z)$;
$\tau$'s act on the Nambu space and $\eta$'s act on the sublattice space.
Adding TRS to this Hamiltonian by using the freedom of the spin space, we obtain
\begin{align}
    \label{eq:DH_chiral_model}
    &\tilde{H}_2(k_x, k_y) \notag\\
    & \quad = (m-t\cos{k_y}) \eta_0\tau_z s_0 \notag\\
    & \quad\quad + \kappa \sin{(k_x/2)} \lbrack \cos{(k_x/2)} \eta_x + \sin{(k_x/2)} \eta_y \rbrack \tau_x s_z \notag\\
    & \quad\quad + \Delta \sin{k_y} \eta_0\tau_y s_0 + 2\alpha \sin{k_y} \eta_z\tau_x s_x.
\end{align}
Here, $s_i~(i=x,y,z)$ are Pauli matrices in the spin space.
When the real number $\alpha$ is zero, the non-Hermitian Hamiltonian is decomposed into Hatano-Nelson models;
$H_2(k_x, k_y)$~[Eq.~(\ref{eq:DH_nu2_model})] is written as
\begin{eqnarray}
    \label{eq:H2alpha0_Bdiag}
    H_2(k_x, k_y) = m\eta_0\rho_0 + \left(\begin{array}{cc}
        H^{(\mathrm{HN})}(k_y)\eta_0 & 0 \\
        0 & H^{(\mathrm{HN})}(-k_y)\eta_0
    \end{array}\right)_\rho, \nonumber \\
\end{eqnarray}
at $k_x=0$.
Here, $H^{(\mathrm{HN})}(k) \equiv -t\cos{k} + i\Delta\sin{k}$ is the Hatano-Nelson model~\cite{Hatano_Nelson_PRL1996, Hatano_Nelson_PRB1997},
which belongs to class A and can exhibit the NHSE characterized by a $\mathbb{Z}$-invariant.
The Hamiltonian $\tilde{H}_2$ belongs to class DIII and satisfies the glide symmetry.
The representation of each symmetry operator is the same as the case of $\tilde{H}_1$.
If we choose the basis such that CS is represented as $\Gamma = \rho_0\chi_z$ where $\rho$'s and $\chi$'s are Pauli matrices, $\tilde{H}_2$ is written as
\begin{equation}
    \label{eq:DH_nu2_model}
    \tilde{H}_2(k_x, k_y) = \left(\begin{array}{cc}
        0 & H_2(k_x, k_y) \\
        H_2^\dagger(k_x, k_y) & 0
    \end{array}\right)_\chi.
\end{equation}
This Hamiltonian is the doubled Hermitian Hamiltonian with $E_\mathrm{ref} = 0$ for the toy model defined in Eq.~(\ref{eq:nu=2_model}).

%%%%%%%%%%%%%%%%%%
\section{Edge modes of the doubled Hermitian Hamiltonian}
\label{app:why_NHSE}
%%%%%%%%%%%%%%%%%%
In Sec.~\ref{sec:Z4inv_sym}, the NHSE is observed at a specific wave number (i.e., $k_x=0$) for the Hamiltonian whose $\mathbb{Z}_4$-invariant takes $\nu = 1, 2$.
This behavior is explained in terms of exact zero modes~\cite{Okuma_TopoOrigin_PRL2020}.

\begin{figure}[htbp]
        \begin{minipage}{0.48\linewidth}
            \includegraphics[width=0.95\linewidth]{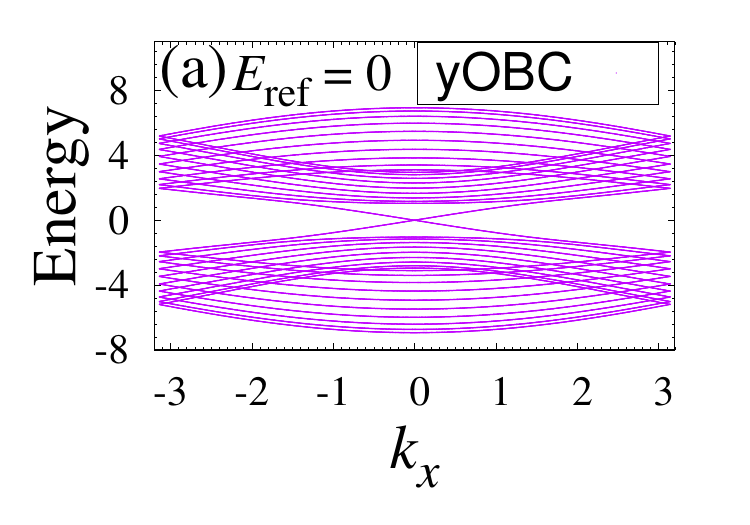}
        \end{minipage}
        \begin{minipage}{0.48\linewidth}
            \includegraphics[width=0.95\linewidth]{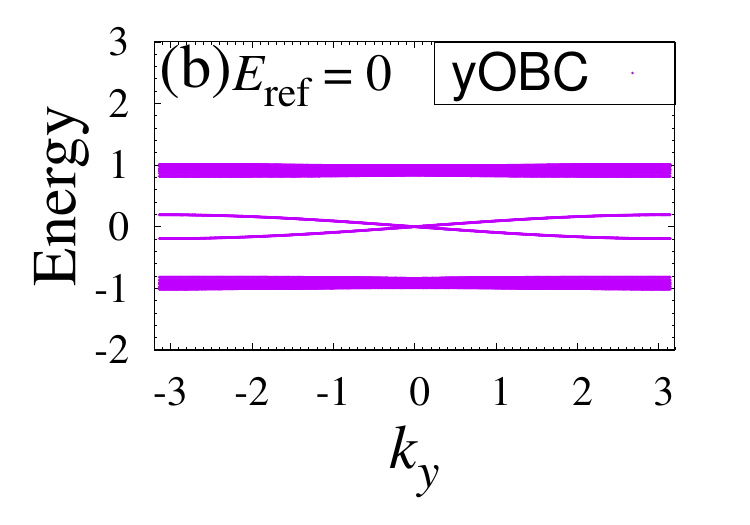}
        \end{minipage}\\
    \caption{ (a): Energy spectra of the doubled Hamiltonian constructed from Eq.~(\ref{eq:nu=1_model}) under yOBC for $(L, m, t, t_{\mathrm{sp}}) = (10, 3, 1, 0.8)$ and $k_x = -\pi + 2\pi n/10^3 (n = 1, \ldots, 10^3)$.
     (b): Energy spectra of the doubled Hamiltonian constructed from Eq.~(\ref{eq:nu=2_model}) under yOBC for $(L, m, t, \kappa, \Delta, \alpha) = (10,0, 1, 0.2, 0.8, 0.1)$ and $k_x = -\pi + 2\pi n/10^3 (n = 1, \ldots, 10^3)$.
    }
    \label{fig:DH_spectra_nu1and2}
\end{figure}

Figure~\ref{fig:DH_spectra_nu1and2} (a) [\ref{fig:DH_spectra_nu1and2} (b)] shows the energy spectrum for the doubled Hermitian Hamiltonian constructed from Eq.~(\ref{eq:nu=1_model}) [(\ref{eq:nu=2_model})] under yOBC.
These edge modes become zero modes at $k_x = 0$.
The same result is obtained when the $E_{\mathrm{ref}}$ equals the eigenenergy of the skin mode.
These results are consistent with the correspondence between the right eigenstates of the non-Hermitian Hamiltonian $H$ and the edge modes of the doubled Hermitian Hamiltonian $\tilde{H}$.
That is, the appearance of $\ket{E_{\mathrm{ref}}}$ satisfying $H\ket{E_{\mathrm{ref}}} = E_{\mathrm{ref}}\ket{E_{\mathrm{ref}}}$ under yOBC implies that the existance of boundary modes $\left( 0, \ket{E_{\mathrm{ref}}} \right)_\chi^T$ satisfying $\tilde{H} \left( 0, \ket{E_{\mathrm{ref}}} \right)_\chi^T = 0$.
The NHSE is not observed for $k_x \neq 0$ because of the absence of zero modes in $\tilde{H}$.

%%%%%%%%%%%%
%%%%%%%%%%%%

\begin{figure}[htbp]
    \begin{minipage}{0.48\linewidth}
        \includegraphics[width=0.95\linewidth]{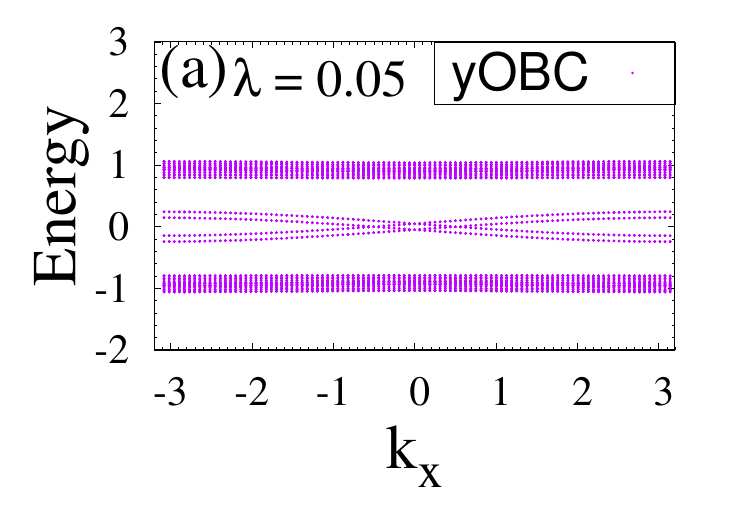}
    \end{minipage}
    \begin{minipage}{0.48\linewidth}
        \includegraphics[width=0.95\linewidth]{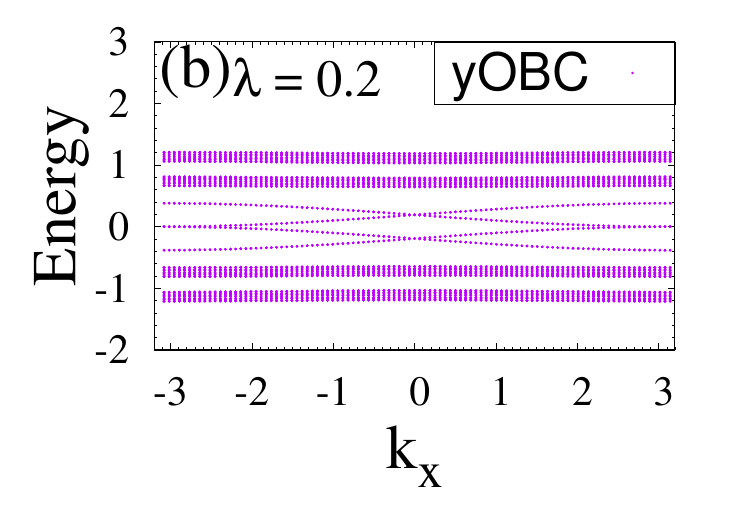}
    \end{minipage}
    \begin{minipage}{0.48\linewidth}
        \includegraphics[width=0.95\linewidth]{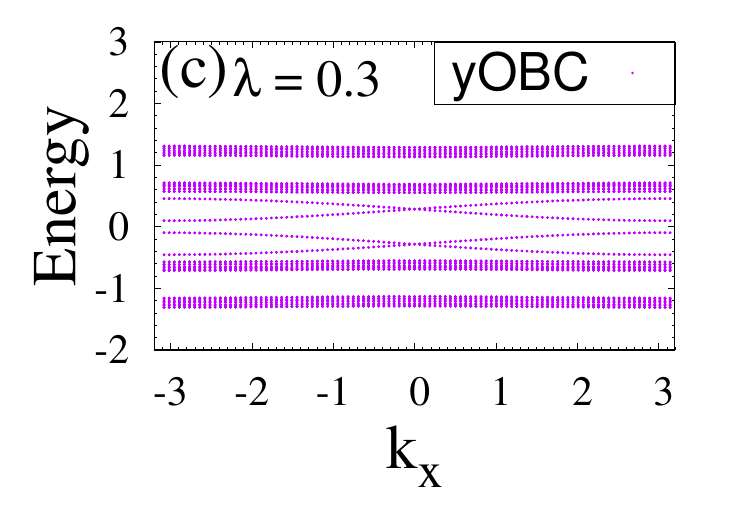}
    \end{minipage}\\
    \caption{(a), (b) and (c): Energy spectra of the doubled Hamiltonian constructed from $H_{4}$ [see Eq.~(\ref{eq:nu=4_model})] under yOBC for $\lambda = 0.05$, $0.2$, and $0.3$.
    These data are obtained for $(L, m, t, \kappa, \Delta, \alpha, E_{\mathrm{ref}}) = (10, 0, 1, 0.2, 0.8, 0.1, 0)$ and $k_x = -\pi + 2\pi n/10^2 (n = 1, \ldots, 10^2)$.
    }
    \label{fig:DH_spectra_nu4}
\end{figure}

The argument of doubled Hermitian Hamiltonian explains the destruction of the NHSE.
Figure~\ref{fig:DH_spectra_nu4} shows the energy spectra of the doubled Hermitian Hamiltonian constructed from Eq.~(\ref{eq:nu=4_model}) under yOBC.
For $\lambda \leq 0.2$, there are zero modes at specific wave numbers [see Figs.~\ref{fig:DH_spectra_nu4}(a) and \ref{fig:DH_spectra_nu4}(b)] which the NHSE for Eq.~(\ref{eq:nu=4_model}) are observed.
On the other hand, for $\lambda > 0.2$ such that $\lambda = 0.3$, the zero modes disappear [see Fig.~\ref{fig:DH_spectra_nu4}(c)].
This result is consistent with the disappearance of the NHSE for Eq.~(\ref{eq:nu=4_model}) when $\lambda > 0.2$.

%%%%%%%%%%%%
%%%%%%%%%%%%

%%%%%%%%%%%%
\section{Topological invariant for class AII${}^\dagger$}
\label{sec:topo_without_glide}
%%%%%%%%%%%%
In section~\ref{subsec:NHSE_nu2}, we consider the Hamiltonian defined in Eq.~(\ref{eq:nu=2_model}) which belongs to the symmetry class AII${}^\dagger$ and has glide symmetry added.
The subsystem at $k_x=0$ \lbrack i.e., $H'_2(k_y) \equiv H_2(0, k_y)$\rbrack belongs to one-dimensional class AII${}^\dagger$ which is characterized by the following $\mathbb{Z}_2$-invariant~\cite{Qi_TopoFermi_PRB2010,Schnyder_TopoFlat_PRB2011,Budich_TopoInv_PRB2013,Kawabata_NHSymTopo_PRX2019} $\theta(E_{\mathrm{ref}}) \in \{0, 1\}$:
\begin{align}
    \label{eq:AIIdagger_Z2_invariant}
    &(-1)^{\theta(E_{\mathrm{ref}})} \notag\\
    & \quad \equiv \mathrm{sgn} \left\{\frac{\mathrm{Pf}\lbrack (H'_2(\pi) - E_{\mathrm{ref}})\mathcal{T} \rbrack}{\mathrm{Pf}\lbrack (H'_2(0) - E_{\mathrm{ref}})\mathcal{T} \rbrack}\right. \notag\\
    & \quad\quad \left.\times \exp\left[-\frac{1}{2}\int_{k=0}^{k=\pi} d\log{\det{\lbrack H'_2(k) - E_{\mathrm{ref}} \rbrack \mathcal{T}}} \rbrack\right] \right\}.
\end{align}
We can compute numerically $\theta(E_{\mathrm{ref}})$ for $H'_2(k_y)$ and confirm that 
$\theta$ takes 0 [$\theta(E_{\mathrm{ref}})=0$] for the parameter values used in Sec.~\ref{subsec:NHSE_nu2} if we set $E_{\mathrm{ref}}$ to the eigenenergy of the skin mode.

This result is understood as follows:
the subsystem at $k_x=0$ can be regarded as two copies of a one-dimensional system with $\theta(E_{\mathrm{ref}})=1$ [see Eq.~(\ref{eq:H2alpha0_Bdiag})].
Thus, in total we have $\theta(E_{\mathrm{ref}})=2=0\, (\mathrm{mod}\, 2)$ for $H'_2(k_y)$.

%%%%%%%%%%%%
\section{Spectrum and eigenstates under xOBC and yPBC}
\label{sec:xOBC_yPBC}
%%%%%%%%%%%%
If we impose the xOBC, the glide symmetry is not closed because the glide operation involves the translation by half a lattice constant in the $x$-direction.
In this case, the NHSE is not observed for Hamiltonian Eq.~(\ref{eq:nu=2_model}).

\begin{figure}[htbp]
    \begin{minipage}{0.48\linewidth}
        \includegraphics[width=0.95\linewidth]{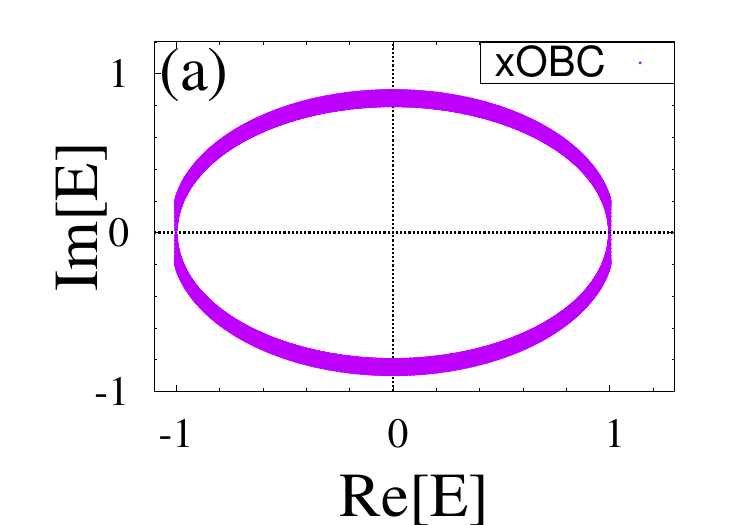}
    \end{minipage}
    \begin{minipage}{0.48\linewidth}
        \includegraphics[width=0.95\linewidth]{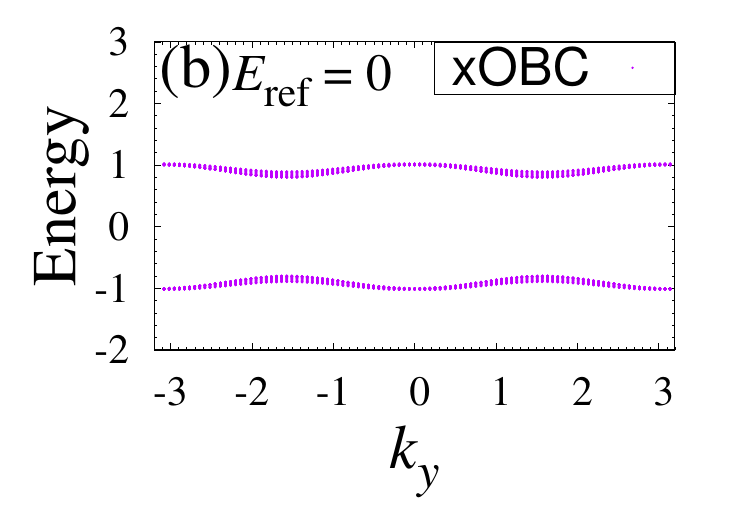}
    \end{minipage}\\
\caption{ (a): Energy spectra of $H_2$ [Eq.~(\ref{eq:nu=2_model})] under xOBC for $k_y = 2\pi n/10^3 (n = 0, \ldots, 10^3-1)$.
 (b): Energy spectra of the doubled Hamiltonian constructed from Eq.~(\ref{eq:nu=2_model}) under xOBC for $k_y = -\pi + 2\pi n/10^2 (n = 1, \ldots, 10^2)$ and $E_{\mathrm{ref}} = 0$.
 These data are obtained for $(L, m, t, \kappa, \Delta, \alpha) = (10, 0, 1.0, 0.2, 0.8, 10^{-1})$.
 }
\label{fig:nu2_xOBC}
\end{figure}

We compute the energy spectrum under xOBC and yPBC [see Fig.~\ref{fig:nu2_xOBC}(a)].
Figure~\ref{fig:nu2_xOBC}(a) shows that the energy spectrum exists away from the real axis, which is similar to the case of xPBC and yPBC.
Thus, the NHSE disappears when the glide symmetry is not closed.
This result implies that the NHSE for $\nu = 2$ is protected by the glide symmetry.
Besides, we confirm that the edge modes do not appear for the doubled Hamiltonian constructed from Eq.~(\ref{eq:nu=2_model}) under xOBC and yPBC [see Fig.~\ref{fig:nu2_xOBC}(b)].

%%%%%%%%%%%%
%%%%%%%%%%%%

\nocite{*}


\begin{thebibliography}{96}
\bibitem [{TRS()}]{TRSB_ptb_ftnt}%
  \BibitemOpen
  \href@noop {} {}\bibinfo {note} {In the presence of $\mathrm{TRS}^\dagger$,
  the skin modes localized around right and left edges are degenerated, which
  yields a numerical difficulty to capture skin modes. Thus, we have introduced
  a perturbation to lift this degeneracy.}\BibitemShut {Stop}%
\end{thebibliography}
\end{document}